\documentstyle[twocolumn,epsfig,seceq]{jpsj}

\def\runtitle{Snapshot Observation for 2D Classical Lattice Models by CTMRG}
\def\runauthor{Kouji {\sc Ueda}, Ryota {\sc Otani}, Yukinobu {\sc Nishio},
Andrej {\sc Gendiar}, and Tomotoshi {\sc Nishino}}

\title{Snapshot Observation for 2D Classical Lattice Models \\
by Corner Transfer Matrix Renormalization Group}

\author
{ Kouji {\sc Ueda}, Ryota {\sc Otani}, Yukinobu {\sc Nishio},
Andrej {\sc Gendiar}$^{2)}$, and Tomotoshi {\sc Nishino}
}

\inst
{Department of Physics, Faculty of Science, Kobe University, Kobe 657-8501\\
$^{2)}$ Institute of Electrical Engineering, Slovak Academy of Sciences, D\'ubravsk\'a cesta 9, 
SK-842 39 Bratislava, Slovakia}

\abst
{
We report a way of obtaining a spin configuration snapshot, which is one of the
representative spin configurations in canonical ensemble, in a finite area of 
infinite size two-dimensional (2D) classical lattice models. The corner 
transfer matrix renormalization group (CTMRG), a variant of the density matrix 
renormalization group (DMRG), is used for the numerical calculation. The matrix 
product structure of the variational state in CTMRG makes it possible to stochastically 
fix spins each by each according to the conditional probability with respect to its environment.
}

\begin{document}
\sloppy
\maketitle

\section{Introduction}

For a classical lattice spin system that interacts with a reservoir,
the spin configuration observed at an instant, which we call {\it snapshot} in
the following, is one of the representative configurations 
in the canonical ensemble. Such a snapshot is experimentally observed if the 
time scale of the observation is much shorter than that of the evolution of the
system. A {\it freezed} spin configuration after sudden cooling can also be 
regarded as a kind of snapshot. In general, snapshots show rough outlook of
the system. For example, the typical size of a spin inverted island in the ordered 
state is of the order of the correlation length, and symmetries of ordered states
can be intuitively identified. 
Figure 1 shows a snapshot inside the area of $100$ by $100$ of 
the two-dimensional (2D) ferromagnetic Ising model, where the system size is 
much larger than the shown region. Normally such a snapshot is drawn by 
Monte Carlo simulation,\cite{Binder} while the shown one is created by the corner 
transfer matrix renormalization group\cite{CTMRG} (CTMRG), a variant of the
density matrix renormalization group\cite{White} (DMRG) applied to
2D classical lattice models. So far both DMRG and CTMRG have been used for 
calculations of thermal average of spin correlation functions, but not for the
snapshots.

In this article we report a way of obtaining the conditional probability for a row 
of spins of length $N$ surrounded by the rest of the system, using the matrix 
product structure\cite{MPS} (MPS) of the variational state employed in 
CTMRG and DMRG. These spins can be fixed one by one 
according to the conditional probability, and after fixing all the spins in the row 
it is possible to obtain the conditional probability for the next row in the same 
manner. Applying such a fixing process for $M$ numbers of rows, one obtains 
a snapshot for the area of $N$ by $M$ in infinite --- or sufficiently large --- 2D 
lattice systems. Numerical cost for this snapshot observation is of the same order 
of conventional zipping process in the finite system DMRG algorithm.\cite{White,DMRG98} 

\begin{figure}
\epsfxsize=65mm 
\centerline{\epsffile{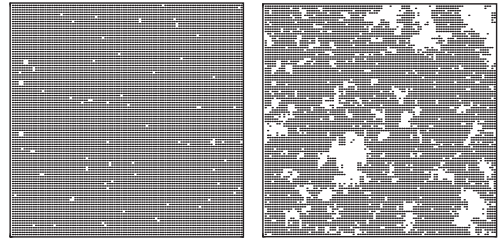}}
\caption{Snapshots of the ferromagnetic Ising model at $T = 1.5$ (left) and $2.27$ (right),
where we have chosen the nearest neighbor coupling constant as the unit of energy.}
\label{fig:1}
\end{figure}

In the next section we briefly explain how to fix a spin in the 2D lattice 
model in terms of the corner transfer matrix formalism. In section 3 we generalize the
spin fixing procedure to a row of spins of the length $N$, by taking partial
sum for the inner product between MPSs. Introducing position dependence to the 
local factors that construct MPS, we successively fix the $M$-rows of spins as we explain 
in Sec.4. In the last section we conclude the result and discuss the relation 
with quantum observation in one dimension.

\section{One Spin Fixing}

The corner transfer matrix (CTM), which was invented by Baxter,\cite{Baxter} is
not only useful for rigorous analyses of 2D lattice systems, but also
efficient for numerical calculations of thermodynamic functions, especially away from
the critical point. Let us briefly look at the construction of CTM and the block spin 
transformation applied for it.

\begin{figure}
\epsfxsize=60mm 
\centerline{\epsffile{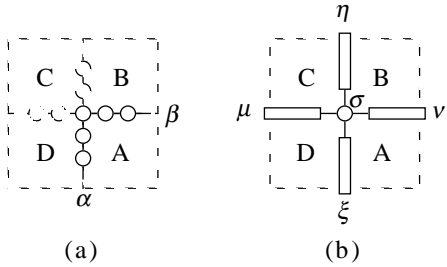}}
\caption{Division of the 2D lattice into 4 corners. (a) Variables of the corner 
transfer matrix A. (b) Renormalized corner transfer matrices.}
\label{fig:2}
\end{figure}

We consider the square lattice Ising model as an example of the 2D classical
lattice spin models. The partition function of the system is expressed as
\begin{equation}
Z = \sum_{\{ \sigma \}}^{~} \exp ( H\{ \sigma \} ) \, ,
\end{equation}
where $\{ \sigma \}$ represents a configuration of Ising spins $\sigma = \pm 1$ 
on the square lattice, and  $H\{ \sigma \}$ is the Ising Hamiltonian, that is
a sum of Ising interaction over all the bonds. 

In the CTM formalism, the 2D lattice is divided into 4 parts, so called the 
quadrants or the corners.\cite{Baxter} Let us label each quadrant as $A$, $B$, $C$,
and $D$. The Hamiltonian $H\{ \sigma \}$ is then expressed as sum of the 
corresponding parts
\begin{equation}
H_A^{~}\{ \sigma_A^{~} \} +
H_B^{~}\{ \sigma_B^{~} \} +
H_C^{~}\{ \sigma_C^{~} \} +
H_D^{~}\{ \sigma_D^{~} \} \, ,
\end{equation}
where
$\{ \sigma_A^{~} \}$,
$\{ \sigma_B^{~} \}$,
$\{ \sigma_C^{~} \}$, and
$\{ \sigma_D^{~} \}$ denote spin configurations in each quadrant
of the system. Note that neighboring quadrants share the same
spins at the boundary between them. In this way the Boltzmann weight of the whole system is 
expressed as a product of 4 factors
\begin{eqnarray}
\exp ( H\{ \sigma \} )   = && \,
\exp ( H_A^{~}\{ \sigma_A^{~} \} )
\exp ( H_B^{~}\{ \sigma_B^{~} \} ) \nonumber\\
~ && \,
\exp ( H_C^{~}\{ \sigma_C^{~} \} )
\exp ( H_D^{~}\{ \sigma_D^{~} \} ) \, .
\end{eqnarray}
The corner transfer matrix is the partial sum of each factor
with respect to the spins {\it inside} the corner
\begin{equation}
A( \alpha \sigma \beta ) = \sum'_{\{ \sigma_A^{~} \}}
\exp ( H_A^{~}\{ \sigma_A^{~} \} ) 
\end{equation}
except for those spins at the boundary with other corners, where the 
notation $\sum'$ denotes this restricted summation, and where $\alpha$ and 
$\beta$ denotes groups of spins at the boundary. (See Fig.~2 (a).) The CTM is 
a block diagonal matrix with respect to $\sigma$,
while in this article we do not use this matrix property explicitly. Other CTMs $B$, $C$, 
and $D$ are defined in the same manner. If the system is
invariant under the rotation of 90 degree these CTMs are equivalent. For 
simplicity, we assume this symmetry in the following.

The matrix dimension of the CTM increases exponentially with the
system size. In order to avoid this blow up and treat the CTM accurately in
numerical calculation, block spin transformation is introduced in CTMRG 
from those spins $\alpha$ and $\beta$ at the boundary to $m$-state auxiliary variables. 
Such a transformation maps the CTM $A$ into the compressed (or renormalized) one
\begin{equation}
A( \xi \sigma \nu ) = \sum_{\alpha \beta}^{~}  \,
U_{\xi}^{~}( \alpha ) \,
U_{\nu}^{~}( \beta  ) \, 
A( \alpha \sigma \beta )  \,  \, ,
\end{equation}
where $\xi$ and $\nu$ are the $m$-state auxiliary variables. The transformation
matrix $U_{\xi}^{~}( \alpha )$ is obtained by diagonalizing the 
density matrix $\rho = ABCD$.\cite{White, Nishino} Other CTMs can also be
mapped to $B( \nu \sigma \eta)$, $C( \eta \sigma \mu )$, and $D( \mu \sigma \xi )$
as shown in Fig.~2 (b). The approximate partition function
\begin{equation}
Z' = \!\!\! \sum_{\sigma^{~} \xi \nu \eta \mu}^{~}
A( \xi \sigma \nu ) \,
B( \nu \sigma \eta) \,
C( \eta \sigma \mu ) \,
D( \mu \sigma \xi )
\end{equation}
is close enough to the original one $Z$ in Eq.~(2.1) if $m$ is sufficiently
large; normally $m$ is of the order of $10$ to $1000$. 

When the system size is far larger than the correlation length of the 
system, the renormalized CTMs becomes independent of the system size
except for a constant multiple. Throughout this article we assume
such a condition, and regard the renormalized CTM as a quadrant of
{\it infinite} size systems. 

\begin{figure}
\epsfxsize=65mm 
\centerline{\epsffile{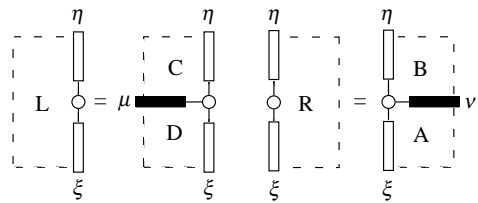}}
\caption{The left and the right vectors created from CTMs.}
\label{fig:3}
\end{figure}

Now we calculate the probability of observing $\sigma = 1$ (up) and
$\sigma = -1$ (down) at the center of the system. We first
prepare two partial sums
\begin{eqnarray}
L( \eta \sigma \xi ) &=& \sum_{\mu}^{~}
C( \eta \sigma \mu ) \,
D( \mu \sigma \xi ) \nonumber\\
R( \xi \sigma \eta ) &=& \sum_{\nu}^{~}
A( \xi \sigma \nu ) \,
B( \nu \sigma \eta) \, ,
\end{eqnarray}
that we regard $2m^2_{~}$-dimensional vectors in the following.
Figure 3 shows the graphical representation of the above equation.
The order of variables in $L( \eta \sigma \xi )$ is opposite to that
of $R( \xi \sigma \eta )$, since we keep the order of auxiliary
variables in the right hand sides of the above equations.
The approximate partition function $Z'$ is represented as
\begin{equation}
Z' = \langle L | R \rangle \equiv \sum_{\xi \sigma \eta}^{~}
L( \eta \sigma \xi ) \, R( \xi \sigma \eta ) \, ,
\end{equation}
where we have introduced bracket notations for the book keeping. 
The probability of $\sigma$ taking the specific value $\bar\sigma$ can be written as
\begin{equation}
p( \bar\sigma ) =
\frac{\langle L  | \, \delta ( \bar\sigma, \sigma ) \,  | R \rangle}{
\langle L  | R \rangle} =
\frac{\sum_{\xi \eta}^{~}
L( \eta \bar\sigma \xi ) \, R( \xi \bar\sigma \eta )}{
\sum_{\xi \sigma\eta}^{~}
L( \eta \sigma \xi ) \, R( \xi \sigma \eta )} \, ,
\end{equation}
where the probability satisfies the normalization $p( 1 ) + p( -1 ) = 1$.
Therefore, creating a random number $x$ in the range  $[ 0, 1 )$ one
can fix $\sigma$ to $\bar\sigma = 1$ if $x < p( 1 )$, otherwise to
$\bar\sigma = -1$.

\section{Snapshot in a row}

Let us generalize the spin fixing procedure to a $N$ numbers of spins in
a row from $\sigma_1^{~}$ to $\sigma_N^{~}$, where $N$ is much 
smaller than the system size. For this purpose we
introduce the half-row transfer matrices (HRTMs),  that are upper and lower
halves of the transfer matrix $T$ to the horizontal direction. Figure 4 shows
the system that we consider in this section. Between the vectors $L$ and
$R$, there are $N - 1$ numbers of transfer matrix
\begin{equation}
T_i^{~} = S_i^{~} P_i^{~}
\end{equation}
from $i = 1$ to $N -1$, where $S_i^{~}$ and $P_i^{~}$ are HRTMs
\begin{eqnarray}
S_i^{~} &=& S( \xi_i^{~} \, \sigma_i^{~} \, \sigma_{i+1}^{~} \, \xi_{i+1}^{~} ) \nonumber\\
P_i^{~} &=& P( \eta_{i+1}^{~} \, \sigma_{i+1}^{~} \, \sigma_i^{~} \, \eta_i^{~} ) \, .
\end{eqnarray}
We have aligned the variables of HRTMs in the clockwise order 
as CTMs. At the moment all the HRTMs are equivalent, though we put 
indices for identification of their positions.

\begin{figure}
\epsfxsize=65mm 
\centerline{\epsffile{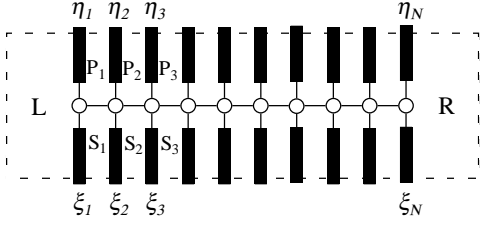}}
\caption{Half-row transfer matrices between vectors $L$ and $R$.}
\label{fig:4}
\end{figure}

We fix these $N$ spins from the left to the right, successively calculating the
conditional probability $p_{\bar\sigma_1^{~} \ldots
\bar\sigma_{i-1}^{~}}^{~}( \bar\sigma_i^{~} )$ for the spin
at $i$-site after fixing those spins in the left
$\bar\sigma_1^{~} , \ldots , \bar\sigma_{i-1}^{~}$. The way
of calculation is essentially the same as operator multiplication to
a given matrix product state\cite{Baxter,MPS,DMRG98} (MPS) in
finite system DMRG algorithm, and it is numerically important to prepare partial 
sum of local factors in advance. We write $R$ as $R_N^{~}$ since it 
contains $\sigma_N^{~}$ as its variable. Let us multiply the transfer 
matrices $T_i^{~}$ one by one to the vector $R_N^{~}$. We obtain
\begin{eqnarray}
R_{N-1}^{~} &=& T_{N-1}^{~} \, R_N^{~}\nonumber\\
R_{N-2}^{~} &=& T_{N-2}^{~} \, R_{N-1}^{~} \, , \,\,\, {\rm etc.,}
\end{eqnarray}
down to $R_1^{~}$. In the numerical calculation we do not possess the
transfer matrix $T_i^{~}$ explicitly, but we apply $S_i^{~}$ and
$P_i^{~}$ part by part as
\begin{eqnarray}
R( \eta_i^{~} \sigma_i^{~} \xi_i^{~} ) =
&& \sum_{\sigma_{i+1}^{~} \xi_{i+1}^{~}}^{~}
S( \xi_i^{~} \sigma_i^{~} \sigma_{i+1}^{~} \xi_{i+1}^{~} ) \\
&&\sum_{\eta_{i+1}^{~}}^{~}
R( \xi_{i+1}^{~} \sigma_{i+1}^{~} \eta_{i+1}^{~} )
P( \eta_{i+1}^{~} \sigma_{i+1}^{~} \sigma_i^{~} \eta_i^{~} ) \, , \nonumber
\end{eqnarray}
where the second sum should be taken first. (See Fig.~5.)

\begin{figure}
\epsfxsize=50mm 
\centerline{\epsffile{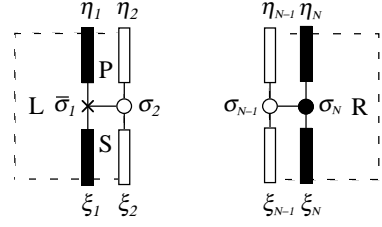}}
\caption{Multiplication of Transfer Matrix.}
\label{fig:5}
\end{figure}

After obtaining $R_1^{~}$ we can calculate the probability for $\sigma_1^{~}$
as before
\begin{equation}
p( \bar\sigma_1^{~} ) = \frac{
\langle L_1^{~}  | \,  \delta ( \bar\sigma_1^{~}, \sigma_1^{~} ) \,  | R_1^{~} \rangle}{
\langle L_1^{~}  | R_1^{~} \rangle} \, ,
\end{equation}
where we have written $L$ as $L_1^{~}$ since it contains $\sigma_1^{~}$ as
its variable. According to this probability we stochastically fix the first spin in the row.
We then consider the conditional
probability $p_{\bar\sigma_1^{~}}^{~}( \bar\sigma_2^{~} )$
for the second spin after fixing the first one. This time, we have to prepare
$L_1^{~}$ multiplied by $T$
\begin{equation}
 L( \eta_2^{~} \sigma_2^{~} \xi_2^{~} ) =
 \sum_{\eta_1^{~}}^{~}
P( \eta_2^{~} \sigma_2^{~} \bar\sigma_1^{~} \eta_1^{~} )
 \sum_{\xi_1^{~}}^{~}
L( \eta_1^{~} \bar\sigma_1^{~} \xi_1^{~} )
S( \xi_1^{~} \bar\sigma_1^{~} \sigma_2^{~} \xi_2^{~} ) 
\end{equation}
as shown in Fig.~5.
 It should be noted that we {\it do not} take configuration sum for
 $\bar\sigma_1^{~}$ since it is already fixed;
$L_2^{~}$ contains a fixed spin $\bar\sigma_1^{~}$ in it.
Using $R_2^{~}$ calculated before and $L_2^{~}$ in Eq.(3.6), we obtain the
conditional probability for the second spin
\begin{equation}
p_{\bar\sigma_1}^{~}( \bar\sigma_2^{~} ) = \frac{
\langle L_2^{~}  | \,  \delta ( \bar\sigma_2^{~}, \sigma_2^{~} ) \,  | R_2^{~} \rangle}{
\langle L_2^{~}  | R_2^{~} \rangle} \, .
\end{equation}
According to this probability we can fix the second spin. After that we 
calculate $L_3^{~}$ in the same manner as Eq.(3.6). Repeating these spin fixing procedure 
to $\sigma_N^{~}$, we finally obtain a spin snapshot $\{ \bar\sigma \} = $
$\bar\sigma_1^{~}$,
$\bar\sigma_2^{~}, \ldots$,
$\bar\sigma_N^{~}$ for a group of $N$ spins in a row.

\section{Snapshot in a Rectangular Area}

\begin{figure}
\epsfxsize=70mm 
\centerline{\epsffile{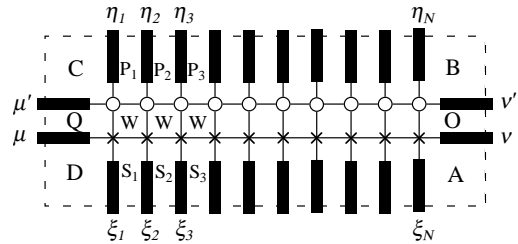}}
\caption{The double-row spin system. 
The cross marks represent the already fixed spins in the 
first row, and the circles are those spins that we will fix
each by each.}
\label{fig:6}
\end{figure}

Let us consider the way of fixing $M$ numbers of spin rows 
successively. The first step is to obtain the probability
for the second spin row under the condition that the first row $\bar\sigma_1^{~}$,
$\bar\sigma_2^{~}, \ldots$, $\bar\sigma_N^{~}$ is already fixed. For the
distinction let us write the spins in the second row as $\{ \tau \} =$
$\tau_1^{~}$, $\tau_2^{~}, \ldots$, $\tau_N^{~}$. Figure 6 shows the
system that we consider for a while, where there is a transfer matrix to the
vertical direction between $\{ \bar\sigma \}$ shown by cross marks and
$\{ \tau \}$ by circles. This transfer matrix consists of the right HRTM 
$O( \nu \bar\sigma_N^{~} \tau_N^{~} \nu' )$, the left one
$Q( \mu' \tau_1^{~} \bar\sigma_1^{~} \mu )$, and $N - 1$ numbers 
of local Boltzmann weights 
$W( \tau_i^{~} \tau_{i+1}^{~} \bar\sigma_i^{~} \bar\sigma_{i+1}{~} )$ 
from $i = 1$ to $N - 1$ in between.~\cite{IRF}

We reduce the above double-row system to the single-row one treated in the previous
section by way of extension of the CTMs and HRTMs to the vertical direction, 
similar to the system size
extension in CTMRG.~\cite{CTMRG} The HRTM $S$ is extended by putting $W$ on top of it
\begin{equation}
S'( \xi_i^{~} \tau_i^{~} \tau_{i+1}^{~} \xi_{i+1}^{~} ) =
W( \tau_i^{~} \tau_{i+1}^{~} \bar\sigma_i^{~} \bar\sigma_{i+1}{~} )
S( \xi_i^{~} \bar\sigma_i^{~} \bar\sigma_{i+1}^{~} \xi_{i+1}^{~} )
\end{equation}
as shown in Fig.~7. Here we do not regard $\bar\sigma_i^{~}$ and $\bar\sigma_{i+1}^{~}$
as the variable of the extended HRTM $S'$ since these are {\it fixed constants}.
Thus the number of elements in $S'_i$ is the same as that of $S_i^{~}$.
It should be noted that after such an extension $S'_i$ becomes position dependent.

\begin{figure}
\epsfxsize=35mm 
\centerline{\epsffile{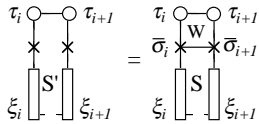}}
\caption{Extension of the HRTM $S$ to the vertical direction.}
\label{fig:7}
\end{figure}

The area extension of CTMs $A$ and $D$ are done by putting HRTMs
$O$ and $Q$
\begin{eqnarray}
A'( \xi_N^{~} \tau_N^{~} \nu' ) &=&
\sum_\nu^{~}
A( \xi_N^{~} \bar\sigma_N^{~} \nu ) \,
O (\nu \bar\sigma_N^{~} \tau_N^{~} \nu' ) \nonumber\\
D'( \mu' \tau_1^{~} \xi_1^{~} ) &=&
\sum_\mu^{~}
Q( \mu' \tau_1^{~} \bar\sigma_1^{~} \mu ) \,
D( \mu \bar\sigma_1^{~} \xi_1^{~} ) \, ,
\end{eqnarray}
respectively, as shown in Fig.~8. 
Again, the number of elements in the extended CTMs are the same as the
original ones. In this way we have reduced the double row system in Fig.~6 as single
row one constructed from $A'$, $B$, $C$, $D'$, $P_i^{~}$, and $S'_i$.
The spin fixing procedure from the left to right explained in the previous
section is now applicable to the second row of spins $\{ \tau \}$.
It is straight forward to repeat the extension of HRTMs and CTMs 
to the vertical direction for $M - 1$ times, we finally obtain the
snapshot of the size $N \times M$ in the infinite system.

\begin{figure}
\epsfxsize=70mm 
\centerline{\epsffile{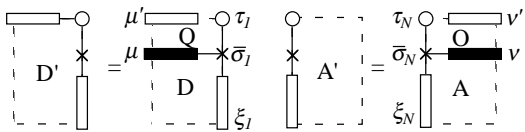}}
\caption{Area extension of CTMs $D$ and $A$.}
\label{fig:8}
\end{figure}

\section{Conclusion and Discussion}

We have explained the way of obtaining spin snapshot in the
area of $N \times M$ for the 2D Ising model.
The snapshot observation can be applied to various 2D lattice models
with short range interaction, such as the IRF and the vertex
models.\cite{water} It is also possible to treat half-infinite or finite
size systems and observe snapshot near the system boundary if we
admit position dependence to HRTMs. (The finite system
DMRG may be more appropriate than CTMRG in such a position dependent
case.)

The calculations of conditional probability requires renormalized CTM
and HRTM, that are converged to the infinite system size limit in the
numerical algorithm of CTMRG. The convergence is quite slow near the
critical temperature, and it is necessary to accelerate it. The same
problem exists in the infinite system DMRG algorithm, and the 
problem has not been solved yet.

It is possible to apply the conventional real space renormalization group 
transformation, such as the majority rule for blocks of spins, to the
obtained snapshots and calculate the critical indices.
This suggests that both CTM and HRTM implicitly have information about
criticality such as critical indices and scaling functions. To pull out
these directly from CTM and HRTM, without looking at the snapshot
is one of the future problem.

We finally comment about a relation between 2D classical systems 
and  1D quantum systems. The snapshot in the former corresponds
to many time observation result in the latter. A snapshot
can be regarded as a representative path in the path integral formulation of  
quantum systems. Successive observations in time-dependent 
DMRG\cite{dynamic1,dynamic2} for
numbers of time slices is the same type of computation.


\begin{thebibliography}{99}
\bibitem{IRF} We have represented the Ising model as the special parameter case
of the IRF model.
\end{thebibliography}
\end{document}